\documentclass{ws-mpla}
\usepackage{color,soul}
\usepackage{hyperref}
\usepackage{mathtools}
\usepackage{url}
\hypersetup{colorlinks=true,urlcolor=cyan,citecolor=green,linkcolor=blue,filecolor=black}
\begin{document}

\catchline{}{}{}{}{}

\title{Neutron lifetime puzzle and Dicke coherence in spontaneous decays}

\author{Zurab K. Silagadze}
\address{Budker Institute of Nuclear Physics and \\ Novosibirsk State University, 630 090, Novosibirsk, Russia. \\ silagadze@inp.nsk.su}

\maketitle

\pub{Received (Day Month Year)}{Revised (Day Month Year)}
\begin{abstract}
The long-standing discrepancy between beam and bottle measurements of the free-neutron lifetime---known as the neutron lifetime puzzle---remains an open anomaly in nuclear and particle physics. Here, we investigate whether collective quantum synchronization via macroscopic Dicke superradiance can influence free-neutron decay or mediate the relaxation of hypothetical excited neutron states. We show that superradiance is fundamentally obstructed in standard three-body weak beta decay by continuous final-state phase space and fermionic antisymmetrization, and remains inactive in terrestrial ultracold neutron traps due to low particle densities, even within the Koch--Hummel excited-neutron model. Extending this analysis to the primordial plasma prior to Big Bang nucleosynthesis, we demonstrate that rapid $e^\pm$ collisional dephasing and intense thermal photon backgrounds prevent a superradiant purge and instead enforce thermal detailed balance. Consequently, across the entire parameter space unconstrained by terrestrial experiments, an excited neutron state inevitably thermalizes before weak freeze-out, increasing the effective neutron statistical weight and severely overproducing primordial $^4\text{He}$, thereby excluding the standard realization of the Koch--Hummel hypothesis.

\end{abstract}

\section{Introduction}
Since the earliest epochs following the Big Bang, neutrons, alongside protons and electrons, have formed the fundamental building blocks of the observable universe. A striking testament to their abundance on a macroscopic scale is the fact that an average human body with a mass of $72~\text{kg}$ contains approximately $40~\text{kg}$ of protons, $32~\text{kg}$ of neutrons, and $22~\text{g}$ of electrons.\cite{Dubbers_2011}

Free neutrons are unstable, and their mean lifetime is a critical parameter across weak interaction physics, nuclear and particle physics, astrophysics, and cosmology.\cite{Dubbers_2011,Abele_2008,Wietfeldt_2011} Since the pioneering measurement of the neutron lifetime in 1951 by Robson\cite{Wietfeldt_2011}, numerous experimental efforts have sought to refine this parameter. A troubling aspect of these collective efforts is their convergence toward two mutually conflicting values. This long-standing discrepancy is widely known as the ``neutron lifetime puzzle''.\cite{Pieter_2018,Fuwa_2024}

These controversial results were obtained via three distinct experimental methodologies: the beam method, the material bottle method, and the magnetic trap method.\cite{Wietfeldt_2011} In the beam method, a flux of slow neutrons passes through a well-defined decay volume, and the charged products of neutron beta decay---protons, electrons, or both---are directly detected.\cite{Wei_2020,Nico_2004,Fuwa_2024} In the bottle method, ultracold neutrons (UCNs) are confined inside a material vessel for storage times comparable to the neutron lifetime, after which the surviving undecayed neutrons are counted.\cite{Mampe_1989,Serebrov_2017} The magnetic trap method is a neutron storage technique in which ultracold neutrons are confined solely by inhomogeneous magnetic fields, thereby eliminating systematic uncertainties associated with material wall interactions during storage.\cite{Paul_1989,Auler_2025}

The most precise experiments utilizing the bottle method yield a world average of $\tau^{\text{bottle}}_n = 878.4 \pm 0.5~\text{s}$, whereas the most accurate measurements employing the beam method yield an average of $\tau^{\text{beam}}_n = 888.0 \pm 2.0~\text{s}$~\cite{Fuwa_2024}. The resulting discrepancy of $4.6\sigma$ ($9.6~\text{s}$) between the two approaches represents the core of the neutron lifetime anomaly.

It should be noted, however, that a recent beam-type experiment at J-PARC\cite{Fuwa_2024} reported a value of $\tau_n = 877.2 \pm 1.7 \, (\text{stat.}) \, {}^{+4.0}_{-3.6} \, (\text{syst.})~\text{s}$. While this result is statistically consistent with the bottle method averages, it exhibits a $2.3\sigma$ tension with prior beam-method measurements. Nonetheless, the systematic uncertainty remains too large to fully resolve the underlying contradiction.

Interestingly, an independent space-based methodology has recently emerged to measure the neutron lifetime. Galactic cosmic rays bombarding planetary surfaces generate high-energy spallation neutrons. A fraction of these neutrons thermalize via collisions with surface and atmospheric material before escaping back into space, where they eventually undergo weak decay. The neutron lifetime can be determined by mapping the flux of remaining neutrons at varying orbital altitudes using satellite-borne neutron detectors.\cite{Lawrence_2021} The technical feasibility of this technique has been demonstrated using legacy data from the Lunar Prospector Neutron Spectrometer,\cite{Wilson_2020,Vydula_2025} though achieving competitive precision will ultimately require a dedicated space-borne neutron lifetime mission.\cite{Vydula_2025}

Alongside experimental refinements, numerous theoretical proposals have been put forward to explain the beam--bottle discrepancy, and we review these proposals in the next section. In the remainder of the manuscript, we investigate whether collective quantum synchronization via Dicke superradiance can influence free-neutron decay and mediate the relaxation of such excited states, beginning with a brief review of Dicke superradiance. We then identify the fundamental quantum-statistical obstructions to superradiance in three-body beta decay and evaluate density and cavity-QED prospects for two-body radiative transitions of putative excited neutrons in terrestrial UCN traps. Finally, we assess the cosmological viability of the Koch--Hummel hypothesis.

\section{Theories Addressing the Neutron Lifetime Puzzle}
``As is well known, discrepancies are a Good Thing, providing a stimulus to further work by both experimentalists and theorists.''\cite{Sucher_1978} It is therefore not surprising that the enigma of the neutron lifetime has stimulated a diverse array of theoretical frameworks seeking its resolution.

A fundamental distinction exists between the two primary measurement methodologies: the storage (bottle) technique measures the total lifetime, whereas the beam method determines the partial lifetime restricted to specific standard beta decay channel. Consequently, a natural solution to this discrepancy would be the existence of an additional, exotic neutron decay channel with a branching fraction of approximately one percent. For instance, a neutron could decay into a hydrogen atom and an antineutrino;\cite{Galzenati_1960,Nemenov_1980} however, the expected branching ratio for this bound-state beta decay channel is far too small to account for the observed lifetime puzzle.\cite{Green_1990} 

A hypothetical strong enhancement of the bound-state beta decay channel as a resolution of the neutron lifetime enigma was postulated in Ref.~\refcite{Oks_2025}. The idea relies on utilizing the second, irregular solution of the Dirac equation, which diverges strongly at the origin ($\propto r^{-1-\sqrt{1-(Z\alpha)^2}} \approx r^{-2}$ for $Z=1$) and is standardly rejected as non-normalizable for $Z\alpha \le \sqrt{3}/2$. In Ref.~\refcite{Oks_2025}, regularization of this singularity is assumed to be provided by the finite size of the proton. However, the existence, normalizability, and physical admissibility of singular solutions to the Dirac equation in a Coulomb field is a classic problem in mathematical physics, investigated in detail within the theory of self-adjoint extensions of differential operators and in the context of superheavy ($Z>137$) nuclei physics.\cite{Case_1950,Zeldovich_1972,Voronov_2006,Gitman_2012} When properly matching the interior and exterior wavefunctions at the proton boundary $R_p$, any realistic finite nuclear potential suppresses the coefficient of the singular exterior component by powers of $R_p/a_0$ (where $a_0$ is the Bohr radius), rendering its contribution negligible. Obtaining the unsuppressed singular solution postulated in Ref.~\refcite{Oks_2025} would require either violating the continuity of the Dirac spinor at the boundary or assuming an unphysical, infinitely deep attractive potential inside the proton --- which would radically distort the experimentally measured Lamb shift and hyperfine structure.

Alternatively, the final state of this missing decay channel could involve one or more dark-sector particles.\cite{Fornal_2018} Nonetheless, many of these proposed dark decay modes face severe constraints from both neutron star physics and precision laboratory experiments.\cite{Klopf_2019} 

A specific realization of dark-sector physics introduces mirror matter as a potential resolution.\cite{Berezhiani_2019} In this framework, an ordinary neutron mixes with its mirror partner, generating a tiny mass splitting of $\Delta m \sim 10^{-7}~\text{eV}$. Consequently, within the several-Tesla magnetic fields utilized in beam experiments, the neutron--mirror-neutron transition undergoes resonant enhancement. This processes converts roughly 1\% of the beam's neutrons into mirror neutrons, which subsequently decay invisibly within the mirror sector. This mechanism elegantly explains why beam experiments, which monitor exclusively ordinary secondary particles, record an artificially prolonged lifetime.\cite{Berezhiani_2019} However, subsequent experimental evidence has placed stringent constraints on this scenario,\cite{Broussard_2022,Gonzalez_2024,Brenot_2026} illustrating what Thomas Henry Huxley famously termed ``the great tragedy of science --- the slaying of a beautiful hypothesis by an ugly fact.''

An alternative variant of dark-sector oscillations is presented in Ref.~\refcite{Dvali_2023}, which examines neutron oscillations into a sterile fermion propagating through large extra dimensions. To date, however, no empirical evidence supporting the existence of large extra dimensions has been observed.\cite{Anchordoqui_2025}

Still another interesting model of this kind can be found in Refs.~\refcite{Tan_2019,Tan_2023}. Contrary to other approaches, this variant of neutron--mirror-neutron oscillations considers the neutron lifetime measured in beam experiments as the true $\beta$-decay lifetime. However, the Standard Model predicts that the neutron lifetime $\tau _{n}$ and its axial-current coupling $g_A=G_A/G_V$ are correlated through the relationship \cite{Czarnecki_2018} 
\begin{equation}
\tau_n(1+3g_A^2)=4908.6\pm 1.9~s,
\label{eq0}
\end{equation}
and the most precise current experimental value of $g_A$ is more consistent with $\tau_n$ from trap experiments than from beam experiments.\cite{Czarnecki_2018}

A compelling observational distinction was highlighted in Ref.~\refcite{Koch_2024} regarding the operational timelines of the two experimental classes. In beam experiments, neutrons are monitored shortly after their extraction from the reactor core, typically on a scale of $2\text{--}25~\text{ms}$. In contrast, bottle experiments measure weak decay rates long after neutron formation ($> 300~\text{s}$). Motivated by this timeline gap, it was postulated that free neutrons are produced with a fraction of long-lived excited states, $n^*$, possessing a modified $\beta$-decay lifetime compared to the ground state. The relaxation of this excited population occurs via gamma-ray emission on a timescale shorter than the typical storage delay of bottle measurements.\cite{Koch_2024} Nevertheless, this excited neutron hypothesis has similarly failed to gain experimental support.\cite{Blatnik_2024,Gonzalez_2026}

Environmental factors have also been explored as a source of the discrepancy. For instance, it has been proposed that the lifetime variation is driven by an enhanced decay probability triggered by multiple elastic collisions of neutrons with each other and with the trap walls.\cite{Desai_2025} This model builds upon foundational studies regarding the influence of matter scattering on particle decay rates.\cite{Koshelkin_1999} Similarly, finite-volume effects originating from the distinct confinement geometries of beam and bottle experiments have been investigated,\cite{Kading_2025} alongside related environmental dependency models detailed in Refs.~\refcite{He_2021,Cea_2021}. Yet, none of these ambient or geometric frameworks have yielded a universally convincing resolution.

Finally, unaccounted-for systematic uncertainties remain a plausible, conventional explanation, particularly within earlier iterations of the beam experiments.\cite{Serebrov_2020} However, these specific technical concerns were rigorously evaluated and subsequently refuted in Ref.~\refcite{Wietfeldt_2023}, reinforcing the baseline accuracy of the beam-method results.

\section{Can Fireflies Provide a Clue?}
We surmise that a remarkable natural phenomenon occurring in the mangrove forests of Southeast Asia may provide a conceptual clue to the neutron lifetime anomaly. A highly vivid historical account of this phenomenon can be found in Ref.~\refcite{Smith_1935}: ``Imagine a tree thirty-five to forty feet high thickly covered with small ovate leaves, apparently with a firefly on every leaf and all the fireflies flashing in perfect unison at the rate of about three times in two seconds, the tree being in complete darkness between the flashes. Imagine a dozen such trees standing close together along the river's edge with synchronously flashing fireflies on every leaf. Imagine a tenth of a mile of river front with an unbroken line of Sonneratia trees with fireflies on every leaf flashing in synchronism, the insects on the trees at the ends of the line acting in perfect unison with those between.''

Following the seminal research on biological rhythms by Winfree\cite{Winfree_1980} (see also the popular exposition in Ref.~\refcite{Strogatz_2003}), the collective flashing of fireflies has become perhaps the most frequently cited illustration of macroscopic synchronization.\cite{Ramirez-Avila_2018} Historically, similar synchronization effects had long been observed in various man-made devices, later finding numerous practical applications in electrical and mechanical engineering.\cite{Pikovsky_2001} Nowadays, synchronization is considered one of the key phenomena of the nonlinear world, with profoundly famous examples in biology.\cite{Landa_1996,Pikovsky_2001,Strogatz_2014} In fact, the phenomenon of synchronization transcends biology, manifesting across a vast array of classical and quantum systems.\cite{Strogatz_1993,Nayfeh_1995,Vicente_2005,Arenas_2008,Gherardini_2018} This raises a compelling theoretical question: could an analogous collective mechanism operate within an ensemble of free neutrons, thereby altering their weak decay dynamics? To date, such collective effects have not been explored in the context of the neutron lifetime puzzle.

\section{Dicke Superradiance}
Under specific environmental conditions --- such as when a collection of atoms occupies a spatial region small compared to the wavelength of the emitted radiation --- the emitters collectively interact through a shared radiation field. This interaction causes the atoms to lock in phase as they decay, culminating in the emission of a short, intense pulse of radiation; a phenomenon known as Dicke superradiance. First predicted by Robert Dicke in 1954,\cite{Dicke_1954} this collective behavior fundamentally differs from standard spontaneous emission characterized by the exponential decay of statistically independent atoms. For comprehensive reviews of the subject, we refer the reader to Refs.~\refcite{Eberly_1971,Gross_1982,Eberly_1987,Ermolaev_1996,Menshikov_1999}.

Assuming that the atomic ensemble is confined to a subwavelength volume, the spatial dependence of the electromagnetic field can be safely neglected. Furthermore, we assume that the atomic ensemble resonantly interacts with only a single mode of the radiation field at the transition frequency 
$\omega$. Under the dipole approximation, the Dicke model Hamiltonian of the system --- which is the multi-atom generalization of the quantum Rabi model --- takes the form:
\begin{align} 
\hat{H} &= \hbar\omega \sum_{\alpha=1}^N \hat{R}^{(\alpha)}_3 + 2g \sum_{\alpha=1}^N \hat{R}^{(\alpha)}_1 \left(\hat{a} + \hat{a}^\dagger\right) + \hbar\omega \left(\hat{a}^\dagger\hat{a} + \frac{1}{2}\right) \nonumber \\ 
&= \hbar\omega \hat{R}_3 + 2g \hat{R}_1 \left(\hat{a} + \hat{a}^\dagger\right) + \hbar\omega \left(\hat{a}^\dagger\hat{a} + \frac{1}{2}\right), 
\label{eq1} 
\end{align} 
where 
\begin{equation} 
\hat{R}^{(\alpha)}_i = \frac{1}{2}\sigma^{(\alpha)}_i, \quad \alpha=1,\ldots,N, \quad i=1,2,3, \qquad \text{and} \qquad \hat{R}_i = \sum_{\alpha=1}^N \hat{R}^{(\alpha)}_i \label{eq2} 
\end{equation} 
denote the individual and collective pseudo-spin operators, respectively. Here, $\hat{a}^\dagger$ and $\hat{a}$ represent the creation and annihilation operators of the resonant field mode, while $g$ represents the atom--field coupling constant, which is directly proportional to the transition dipole moment $\mathbf{d}$.

From the fundamental commutation relations of the Pauli matrices, it follows that the total pseudo-spin squared operator, $\hat{R}^2 = \hat{R}_1^2 + \hat{R}_2^2 + \hat{R}_3^2$, commutes with all individual and collective pseudo-spin components, and consequently, with the total Hamiltonian $\hat{H}$. Therefore, the quantum state of the atomic ensemble $|J,M\rangle$ can be fully characterized by the total pseudo-spin quantum number $J$, giving the eigenvalue of $\hat{R}^2$ as $J(J+1)$, and by the inversion quantum number $M = \frac{1}{2}(N_e - N_g)$, which acts as the eigenvalue of $\hat{R}_3$. Here, $N_e$ and $N_g$ represent the number of atoms populating the excited and ground states, respectively.

Radiative transitions are strictly permitted only between adjacent $|J,M\rangle$ states. Specifically, for the downward transition $|J,M\rangle \to |J,M-1\rangle$, the matrix element of the collective lowering operator $\hat{R}_- = \hat{R}_1 - i\hat{R}_2$ is given by:
\begin{equation}
\langle J,M-1| \hat{R}_- |J,M\rangle = \sqrt{(J+M)(J-M+1)}.
\label{eq3}
\end{equation}
Consequently, under the assumption that the radiative decay cascade initiates from the completely inverted state $|J,\frac{N}{2}\rangle$ (where $J = N/2$), the corresponding collective decay rate is expressed as:
\begin{equation}
\gamma_{M,M-1} = \gamma \left(\frac{N}{2}+M\right)\left(\frac{N}{2}-M+1\right),
\label{eq4}
\end{equation}
where
\begin{equation}
\gamma = \frac{4\omega^3}{3\hbar c^3} \left| \langle g| \hat{\mathbf{d}} |e\rangle \right|^2
\label{eq5}
\end{equation}
is the standard spontaneous decay rate of an isolated individual atom via an electric dipole transition.

At the onset of the decay cascade ($M \approx N/2$), the initial emission rate scales strictly linearly with $N$. Macroscopic coherence develops only as the cascade progresses down the state ladder, driven by the mutual, self-induced coupling of the atomic ensemble to the common electromagnetic mode.

Let $t(N_g)$ be the time elapsed when $N_g$ atoms have transitioned to the ground state from an initially fully inverted state, implying that exactly $N_g$ photons have been radiated by this moment. We can approximate $t(N_g)$ by performing a discrete summation over the inverse transition rates $\gamma^{-1}_{M,M-1}$, which represent the physical lifetimes of the individual step-wise cascade transitions\cite{Ermolaev_1996}:
\begin{equation}
    t(N_g) = \sum_{M=\frac{N}{2}-N_g+1}^{\frac{N}{2}} \gamma^{-1}_{M,M-1} = \sum_{M=\frac{N}{2}-N_g+1}^{\frac{N}{2}} \frac{\tau_0}{\left(\frac{N}{2}+M\right)\left(\frac{N}{2}-M+1\right)},
\label{eq6}
\end{equation}
where $\tau_0 = \gamma^{-1}$ is the lifetime of a single isolated atom. This sum can be evaluated analytically in the large-$N$ limit ($N \gg 1$) using the Euler--Maclaurin formula. Solving for $N_g$, one obtains\cite{Ermolaev_1996} (for pedagogical details, see also Ref.~\refcite{Alizzi_2026}):
\begin{equation}
N_g = N \, \frac{e^{\frac{N}{\tau_0}(t-t_0)}}{e^{\frac{N}{\tau_0}(t-t_0)} + 1}, \qquad \text{where} \qquad t_0 = \frac{\ln N}{N} \, \tau_0.
\label{eq7}
\end{equation}
Consequently, the inversion $M(t)$ evolves as:
\begin{equation}
M(t) = \frac{N}{2} - N_g = -\frac{N}{2} \, \tanh \left[ \frac{N}{2\tau_0}(t-t_0) \right].
\label{eq8}
\end{equation}
The dimensionless energy $W$ (in units of $\hbar\omega$) stored within the collective atomic system is  $W = \frac{1}{2}(N_e - N_g) = M$. Therefore, the total radiated intensity is given by the rate of collective energy loss:
\begin{equation}
I(t) = -\frac{d}{dt} (\hbar\omega M) = \frac{N^2 \hbar\omega}{4\tau_0} \, \operatorname{sech}^2 \left[ \frac{N}{2\tau_0}(t-t_0) \right]=\frac{\hbar\omega}{\tau_0} \,\left(\frac{N}{2}+W\right)\left(\frac{N}{2}-W\right).
\label{eq9}
\end{equation}
As demonstrated by Eq.~\eqref{eq9}, the radiation profile is highly non-monotonic; following a characteristic quantum initiation delay time $t_0$, the ensemble exhibits a macroscopic intensity peak. The collective pulse duration $\tau = \tau_0 / N$ compresses as $1/N$, while the peak coherent intensity scales quadratically as $N^2$. These features represent the classic, defining hallmarks of Dicke superradiance.

\section{Dicke Superradiance in Extended Systems}
In practice, the physical conditions required for ideal Dicke superradiance in subwavelength samples are rarely found in nature and remain difficult to engineer. Nevertheless, Dicke superradiance can occur even in macroscopically extended media.\cite{Dicke_1954} For our purposes, it is sufficient to utilize a phenomenological interpolation formula that bridges the quantum incoherent exponential decay law and the 
coherent radiation pulse of Eq.~\eqref{eq9}, effectively characterizing the superradiant dynamics within an extended sample.\cite{Eberly_1972}

Under the standard exponential decay law $N_e = N e^{-t/\tau_0}$, the dimensionless stored energy $W$ (or macroscopic inversion $M$) has the form
\begin{equation}
W = M = \frac{1}{2}(N_e - N_g) = N_e - \frac{N}{2} = N e^{-t/\tau_0} - \frac{N}{2}.
\label{eq10}
\end{equation}
Differentiating Eq.~\eqref{eq10} with respect to time yields the linear relationship:
\begin{equation}
-\frac{dW}{dt} = \frac{1}{\tau_0} N e^{-t/\tau_0} = \frac{1}{\tau_0} \left( \frac{N}{2} + W \right).
\label{eq11}
\end{equation}
To connect this incoherent regime with the classical coherent decay law described in 
Eq.~\eqref{eq9}, we introduce the nonlinear interpolation formula proposed by 
Eberly~\cite{Eberly_1972}:
\begin{equation}
-\frac{dW}{dt} = \frac{\mu}{\tau_0} \left( \frac{N}{2} + W \right) \left( \frac{N}{2} - W + \frac{1}{\mu} \right),
\label{eq12}
\end{equation}
where $\mu$ denotes the geometric shape factor of the extended sample. 

This general interpolating formula successfully captures both physical limits of the system. Indeed, in initial stage, when the system is fully inverted ($W \approx N/2$), the condition $N/2 - W \ll 1/\mu$ holds. The final term inside the parentheses reduces to $1/\mu$, which cancels out the leading factor of $\mu$, causing Eq.~\eqref{eq12} to reduce directly to the linear exponential decay law of Eq.~\eqref{eq11}. 

In coherent stage, as the cascade progresses ($W \approx 0$), a macroscopic dipole moment builds up, satisfying $N/2 - W \gg 1/\mu$. Under this condition, the $1/\mu$ term becomes completely negligible, and Eq.~\eqref{eq12} cleanly transitions into the quadratic, coherent radiation law of Eq.~\eqref{eq9}, where the peak emission intensity scales as $\mu N^2$.

Letting $t_0$ denote the instant when the collective emission rate $-dW/dt$ in Eq.~\eqref{eq12} reaches its peak value at $W_0 = \frac{1}{2\mu}$, the analytical solution to this differential equation is given by:
\begin{equation}
W(t) = \frac{N}{2} \left[ -\left(1 + \frac{1}{N\mu}\right) \tanh \left( \frac{N\mu + 1}{2\tau_0} (t - t_0) \right) + \frac{1}{N\mu} \right],
\label{eq13}
\end{equation}
which yields a total radiated profile of the form:\cite{Eberly_1972}
\begin{equation}
I(t) = -\hbar\omega \frac{dW}{dt} = \hbar\omega \frac{(N\mu + 1)^2}{4\mu\tau_0} \operatorname{sech}^2 \left( \frac{N\mu + 1}{2\tau_0} (t - t_0) \right).
\label{eq14}
\end{equation}
The dimensionless product $N\mu$ serves as the characteristic parameter governing the 
superradiant dynamics in extended media, effectively playing a role analogous to the total particle number $N$ in localized, subwavelength configurations. Here, the geometric shape factor $\mu$ represents a parameter determined by the specific boundary dimensions and spatial density configuration of the resonant active volume.\cite{Eberly_1971} 

A rigorous quantum-mechanical justification for this interpolation model and its associated dynamics can be found in Ref.~\refcite{Eberly_1971}. For a detailed theory of Dicke superradiance in extended samples, we refer the reader again to Refs.~\refcite{Gross_1982,Eberly_1987,Ermolaev_1996,Menshikov_1999}. Following its theoretical prediction, Dicke superradiance was first experimentally confirmed in optically pumped molecular gases\cite{Skribanowitz_1973} and subsequently demonstrated in atomic vapors\cite{Gibbs_1977} and Bose-Einstein condensates\cite{Inouye_1999,Baumann_2010}. Alongside multi-photon bursts, the weak-excitation regime of single-photon superradiance, governed by so called timed Dicke states,\cite{Scully_2006} has been prominently realized in nuclear ensembles\cite{Rohlsberger_2010} and dilute cold-atom clouds.\cite{Araujo_2016,Roof_2016} Over the past two decades, advances in quantum control have also extended superradiant phenomena to solid-state platforms, including semiconductor quantum dots,\cite{Scheibner_2007} superconducting qubits,\cite{Mlynek_2014} and diamond nitrogen-vacancy ensembles\cite{Angerer_2018} (see Ref.~\refcite{Cong_2016} for a comprehensive review). More recently, research has explored the universality of superradiant bursts in ordered atomic arrays,\cite{Masson_2022} equilibrium magnonic superradiant phase transitions,\cite{Kim_2025} and astrophysical realizations of the phenomenon,\cite{Rajabi_2016-I,Rajabi_2016-II,Rajabi_2020,Alizzi_2026} while sparking debate over the fundamental role of quantum entanglement—or lack thereof—during collective decay.\cite{Bassler_2025} The application of superradiance to weakly bound nuclei, giant resonances, and exotic baryons was considered in Ref.~\refcite{Zelevinsky_2004}, and a very interesting hydrodynamic analog of this important quantum optics effect was demonstrated in Refs.~\cite{Papatryfonos_2022,Frumkin_2021}. As we see, the phenomenon of superradiance is very general.

\section{Dicke coherence in neutron lifetime experiments?}
The application of macroscopic Dicke coherence outside quantum optics is rare. An important recent exception is Ref.~\refcite{Arvanitaki_2024}. In this paper, it was proposed to use superradiant interactions of collective Dicke spin states to increase the extremely small neutrino--nucleus scattering rates by a factor of $N^2$, which would potentially allow the detection of the cosmic neutrino background. This analysis was further extended in Ref.~\refcite{delCastillo_2025} by accounting for realistic noise sources that can degrade the collective signal. Unfortunately, these latter results indicate that the direct detection of the cosmic neutrino background remains beyond the reach of current technology.

An interesting variant of macroscopic Dicke coherence in neutrino physics was the proposed concept of a ``neutrino laser'' via collective nuclear decay. Because neutrinos are fermions and cannot undergo standard stimulated emission, Jones and Formaggio in Ref.~\refcite{Jones_2024} suggested bypassing the Pauli Exclusion Principle by utilizing a dense, radioactive Bose-Einstein condensate of $^{83}\text{Rb}$. They argued that collective electron-capture transitions would achieve an $N^2$ superradiant enhancement, drastically accelerating the nuclear decay rate. However, subsequent theoretical work by several authors\cite{Lin_2025,Lu_2025} rigorously demonstrated the fundamental impossibility of such a device. Due to the anti-commuting nature of fermionic fields, trying to emit multiple neutrinos into 
the same phase-space volume triggers a strict Pauli blockade. This anti-symmetrization leads to destructive quantum interference that forces the emission rate to scale linearly with $N$, rather than quadratically, completely extinguishing any superradiant gain.

Based on the same underlying principles, it becomes clear why putative Dicke coherence in the weak decay of free neutrons cannot be used to resolve the neutron lifetime puzzle. Beyond the fundamental kinematic blockades described above, neutrons are prevented from establishing collective behavior through a shared neutrino field because beta decay is a three-body process. This multi-particle final state produces neutrinos across a continuous spectrum, which inherently precludes the precise phase coupling and subsequent resonance enhancement required to interact with neighboring neutrons. Furthermore, the external magnetic fields employed to trap neutrons in bottle experiments are completely ineffective at inducing coherence within their weak decay channel, owing to the tremendous disparity between the nano-electronvolt trapping scales and the mega-electronvolt scale of the decay.
 
Despite Dicke coherence being ineffective for directly solving the neutron lifetime anomaly, it could be valuable indirectly as an additional ingredient in the excited neutron hypothesis proposed by Koch and Hummel\cite{Koch_2024}, at first sight rendering the model more natural and less fine-tuned. This fine-tuning challenge stems from the requirement that the long-lived excited state, $n^*$, which undergoes a discrete de-excitation channel, must possess a characteristic lifetime $\tau_\gamma$ falling within a narrow window. Specifically, it must remain long enough to survive the flight time of cold neutron beams, yet short enough to ensure that the excited population completely relaxes to the ground state during the storage phase of ultra-cold neutron bottle experiments. Without a dynamic regulatory mechanism, such a stringent environmental constraint requires an unnaturally precise choice for the mechanism governing the transition.

It is tempting to think that Dicke superradiance could significantly alleviate this problem by demonstrating how the kinematic conditions of different experimental setups dictate the onset of coherence. Namely, in beam experiments, neutrons possess relatively high velocities and a large velocity dispersion; the resulting severe Doppler broadening effectively scrambles the quantum phases, destroying any collective behavior and forcing the excited states to decay via a slow, linear regime. Conversely, in bottle experiments, the neutrons are ultra-cold with exceptionally low velocities and negligible dispersion. The absence of Doppler broadening allows the neighboring neutrons to maintain phase-matching, triggering a fast superradiant burst that rapidly relaxes the population to the ground state even if the electromagnetic de-excitation channel's lifetime exceeds 1000 seconds.

However, this collective mechanism imposes a strict constraint on the energy scale of the de-excitation transition. For macroscopic coherence to develop, the wavelength $\lambda$ of the emitted quantum must encompass a large number of interacting particles, requiring $\lambda$ to be comparable to or greater than the average inter-neutron distance. Given typical ultra-cold neutron storage densities of order $10\text{--}100\text{ cm}^{-3}$, the corresponding wavelength must reside at least in the millimeter-to-centimeter range, pointing to a remarkably low transition energy scale $\Delta E \sim 10^{-6}\text{--}10^{-3}\text{ eV}$. Such an exceptionally small mass splitting is profoundly unnatural within the framework of quantum chromodynamics (QCD) for an isolated nucleon in vacuum, where typical hadronic scales reside in the $\text{MeV}$-to-$\text{GeV}$ range, corresponding to a hadron size of about $1\text{--}2\text{ fm}$. Even in the Koch-Hummel framework\cite{Koch_2024}, with a toy quark model explaining how a small energy gap can arise, such a small energy scale seems problematic (unless hypothetical dark sector physics is assumed), since it would imply  astronomically large electromagnetic transition times that cannot be overcome even by Dicke time-reduction.  

Therefore, we choose to assume $\Delta E \sim 1\text{ eV}$ and $\tau_\gamma \sim 100\text{ s}$, values that are not excluded by current experimental constraints\cite{Blatnik_2024}. To trigger the Dicke mechanism under these parameters, the required neutron density proves completely prohibitive. For a small, isotropic sample where coherent coupling is confined within a volume of $\lambda^3$, the density must satisfy $n \gtrsim 10\,\lambda^{-3}$, which demands an exceptionally high value of $n \sim 5.2 \times 10^{12}~\text{cm}^{-3}$. For large, elongated pencil-shaped samples with a length of $L = 100~\text{cm}$ (leaving aside the difficulties of their preparation), the superfluorescence regime is governed by the resonant optical depth $b_0 \propto n \lambda^2 L$~\cite{Ermolaev_1996,Kaiser_2016}. Even in this geometrically favorable case, merely reaching the threshold for collective behavior ($b_0 \sim 10$) requires a minimum density of $n_{\text{min}} \approx 10^7~\text{cm}^{-3}$. Because actual ultracold neutron bottle experiments operate at densities of only $10\text{--}100~\text{cm}^{-3}$, the collective coupling is suppressed by at least five to six orders of magnitude below this threshold, demonstrating the irrelevance of superradiance in this framework.

To overcome the density-dependent limitations of the Dicke mechanism and provide a definitive experimental test of the Koch--Hummel hypothesis, one might instead consider utilizing an electromagnetic resonator. Placing ultracold neutrons within a cavity resonant with the transition frequency ($\lambda \approx 1.24~\mu\text{m}$ for $\Delta E \sim 1~\text{eV}$) modifies the local density of electromagnetic states via the Purcell effect~\cite{Purcell_1946}. In this scenario, the enhancement of the spontaneous decay rate does not rely on collective inter-particle interactions, but rather on the coupling of individual neutrons to the resonant cavity mode. For a vacuum cavity ($n_{\text{refr}} = 1$), the maximum spontaneous emission rate is enhanced by the Purcell factor, defined as
\begin{equation}
F_P = \frac{3}{4\pi^2}\, \lambda^3 \, \frac{Q}{V},
\end{equation}
where $Q$ is the quality factor of the resonator and $V$ is the effective optical mode volume. For a high-$Q$ optical microcavity, this geometric enhancement could in principle compress a baseline radiative lifetime of $\tau_\gamma \sim 100~\text{s}$ down to a fraction of a second, entirely independent of the ultracold neutron storage density. Such an optical intervention would selectively accelerate the relaxation of the $n^*$ population during the storage phase, offering a controlled, density-independent method to test the excited-neutron hypothesis.

However, a closer inspection of the experimental feasibility of this approach reveals a severe technological challenge. While microcavities can achieve a high $F_P$ by compressing $V$ to the scale of $\lambda^3$, they are physically incapable of storing a macroscopic population of ultracold neutrons. Conversely, if we consider a minimally sized macroscopic storage cavity with a volume of just $V = 1~\text{cm}^3$, achieving an enhancement of $F_P \sim 100$ formally requires a quality factor of $Q \approx 6.9 \times 10^{14}$. Not only does this exceed the state-of-the-art limit for optical and near-infrared resonators ($Q \sim 10^{10}\text{--}10^{11}$) by several orders of magnitude, but the standard Purcell formula also breaks down in this regime: for ultracold neutrons moving at typical velocities of $v \sim 5~\text{m/s}$, Doppler broadening ($\Delta\nu_D \sim v/\lambda \approx 4~\text{MHz}$) exceeds the cavity linewidth once $Q \gtrsim 6 \times 10^7$, capping the effective enhancement well below the required threshold. Consequently, a simple cavity intervention is severely bottlenecked by both the spatial mismatch between macroscopic neutron storage needs and microscopic optical mode volumes and the motional broadening of the stored neutrons.

Periodic dielectric structures, such as a three-dimensional photonic crystal or an optical metamaterial engineered directly throughout the ultracold neutron storage volume, could in principle bypass the volume-versus-$Q$ bottleneck of single macroscopic cavities. However, they introduce a severe, if not fatal, experimental complication: material-induced UCN losses. Any realistic solid substrate used to fabricate these structures will inevitably induce neutron losses through radiative nuclear capture and inelastic up-scattering at surface or bulk boundaries, causing rapid depletion of the stored neutron population over sub-micron geometric confinement scales. Consequently, these material-induced loss mechanisms effectively invalidate distributed optical structures as a viable testing ground, indicating that verifying or ruling out the Koch--Hummel hypothesis via direct electromagnetic cavity engineering is practically impossible.

Finally, it is instructive to evaluate the behavior of the excited neutron state $n^*$ in the early universe, specifically at the onset of the Big Bang Nucleosynthesis (BBN) epoch around $t \sim 1~\text{s}$ ($T \sim 1~\text{MeV}$). At this temperature, the ambient blackbody photon density is $n_\gamma \approx 3.17 \times 10^{31}~\text{cm}^{-3}$, and the local baryon-to-photon ratio $\eta_{1\,\text{MeV}} \approx 6.14 \times 10^{-10}$ (consistent with the present-day Planck value)\cite{Aghanim_2020} yields a baryon density of $n_b \approx 1.95 \times 10^{22}~\text{cm}^{-3}$. For a transition energy of $\Delta E \sim 1~\text{eV}$ ($\lambda \approx 1.24 \times 10^{-4}~\text{cm}$), the laboratory Dicke density threshold ($n \gtrsim 10\lambda^{-3} \sim 5.2 \times 10^{12}~\text{cm}^{-3}$) is exceeded by nearly ten orders of magnitude, placing $N \sim n_b \lambda^3 \approx 3.72 \times 10^{10}$ nucleons within a single coherent volume. Although primordial magnetic fields are spatially uniform across $\lambda$---well below the plasma's ohmic damping scale of $\sim 10~\text{cm}$---\cite{Banerjee_2004,Subramanian_2015} and cannot prevent the buildup of coherence via Zeeman dephasing, an idealized superradiant burst with characteristic timescale $\tau_R \sim \tau_\gamma / N \approx 2.69 \times 10^{-9}~\text{s}$ (for $\tau_\gamma \sim 100~\text{s}$) still cannot develop in the primordial plasma.

Two fundamental physical mechanisms prevent Dicke superradiance from operating or acting as a one-way sink to depopulate $n^*$ at $T \sim 1~\text{MeV}$:
 
\emph{Collisional dephasing:} At $T \sim 1~\text{MeV}$, relativistic electron--positron pairs are in thermal equilibrium with photons, giving a lepton number density $n_{e^\pm} \approx \frac{3}{2} n_\gamma \approx 4.8 \times 10^{31}~\text{cm}^{-3}$ (from the Fermi--Dirac factor of $3/4$ and four spin states). Electromagnetic scattering of neutrons off this dense $e^\pm$ bath---mediated by the neutron's anomalous magnetic dipole moment $\mu_n \approx -1.91\,\mu_N$ with characteristic cross section $\sigma_{n e^\pm} \sim \alpha \mu_n^2 \sim 10^{-30}\text{--}10^{-29}~\text{cm}^2$---yields a collision rate $\Gamma_{\text{coll}} \sim n_{e^\pm} \sigma_{n e^\pm} c \sim 10^{12}\text{--}10^{13}~\text{s}^{-1}$. Because $\Gamma_{\text{coll}} \gg \tau_R^{-1} \approx 3.7 \times 10^8~\text{s}^{-1}$, frequent scattering destroys inter-particle phase coherence long before collective emission can build up.

\emph{Thermal detailed balance:} Because $T \sim 1~\text{MeV} \gg \Delta E \sim 1~\text{eV}$, the ambient Bose--Einstein photon occupation number at the transition frequency is enormous, $\bar{n}_{\text{th}} = (e^{\Delta E/T} - 1)^{-1} \approx T/\Delta E \sim 10^6$. Consequently, for $\tau_\gamma \sim 100~\text{s}$, both upward thermal photo-excitation ($\Gamma_{\text{abs}} = (g_{n^*}/g_n)\bar{n}_{\text{th}}\,\tau_\gamma^{-1} \sim 10^4~\text{s}^{-1}$) and downward stimulated emission ($\Gamma_{\text{stim}} = \bar{n}_{\text{th}}\,\tau_\gamma^{-1} \sim 10^4~\text{s}^{-1}$) exceed the radiation-dominated Hubble expansion rate $H = (2t)^{-1} \sim 0.5~\text{s}^{-1}$ at $t \sim 1~\text{s}$ by four orders of magnitude, rapidly locking the $n$--$n^*$ system into thermal equilibrium. Here, $g_n = 2J_n + 1 = 2$ and $g_{n^*} = 2J_{n^*} + 1$ denote the spin-degeneracy factors of the ground-state ($J_n = 1/2$) and excited ($J_{n^*}$) neutron states, respectively. In the three-quark harmonic-oscillator toy model of Koch and Hummel,\cite{Koch_2024} the excited states carry orbital excitations ($l > 0$) whose total angular momentum $J_{n^*} = l \pm 1/2$ yields $g_{n^*} = 2l + 2$ or $2l$ (for instance, $g_{n^*} = 4$ for $J_{n^*} = 3/2$ in the $l = 1$ orbital mode, or higher degeneracies $g_{n^*} \ge 6$ for multi-quantum cascade states), so that $g_{n^*}/g_n \ge 1$.

Consequently, radiative transitions rapidly establish thermal equilibrium with spin-weighted populations ($n_{n^*} / n_n = (g_{n^*}/g_n) e^{-\Delta E/T} \approx g_{n^*}/g_n \ge 1$) prior to weak freeze-out at $T_f \approx 0.8~\text{MeV}$ ($t_f \approx 1~\text{s}$). Because weak interactions ($p + e^- \leftrightarrow n + \nu_e$) maintain chemical equilibrium between protons and ground-state neutrons ($\mu_n = \mu_p$) down to $T_f$, the ground-state ratio remains fixed at $(n_n/n_p)_f = e^{-Q/T_f} \approx 0.198$ (where $Q = m_n - m_p = 1.293~\text{MeV}$); populating the $n^*$ states therefore draws additional baryons from the proton reservoir, boosting the total neutron-to-proton ratio at freeze-out to $(n_n + n_{n^*})/n_p = e^{-Q/T_f}\,[1 + (g_{n^*}/g_n)e^{-\Delta E/T_f}] \approx 0.198\,(1 + g_{n^*}/g_n)$. Moreover, the direct experimental bound $\tau_\gamma < 139~\text{s}$ from the UCN$\tau$ storage search\cite{Blatnik_2024} ensures that radiative transition rates ($\Gamma_{\text{abs}}, \Gamma_{\text{stim}} \gtrsim \tau_\gamma^{-1} > 7.2 \times 10^{-3}~\text{s}^{-1}$) exceed the Hubble expansion rate $H = (2t)^{-1} \gtrsim 2.8 \times 10^{-3}~\text{s}^{-1}$ throughout the entire $\Delta t \approx 180~\text{s}$ cooling interval down to deuterium synthesis ($T_{\text{BBN}} \approx 80~\text{keV}$). Fast radiative equilibration therefore maintains a ground-state fraction $f_n(T) = [1 + (g_{n^*}/g_n)e^{-\Delta E/T}]^{-1}$ all the way to $^4\text{He}$ formation; since excited states are effectively $\beta$-stable ($\tau_\beta^* \gg \tau_\beta \approx 878~\text{s}$), the effective decay rate of the total neutron pool is reduced to $\Gamma_{\beta,\text{eff}} = f_n(T)\,\tau_\beta^{-1}$. Even at the upper experimental limit $\Delta E = 30~\text{keV}$---set by gravitational quantum-state measurements in the qBounce experiment---\cite{Koch_2024} we still have $\Delta E < T_{\text{BBN}}$, raising the neutron survival factor $S_{\text{eff}} = \exp[-\int_{t_f}^{t_{\text{BBN}}} f_n(T(t))\,\tau_\beta^{-1}\,dt]$ from the standard BBN value $S_{\text{std}} = e^{-\Delta t/\tau_\beta} \approx 0.815$ to $S_{\text{eff}} \approx 0.89\text{--}0.95$ (for $g_{n^*} \ge 2$). Since virtually all surviving neutrons at $T_{\text{BBN}}$ are bound into $^4\text{He}$ nuclei (via $n^* \to n + \gamma$ followed by $p + n \to d + \gamma$), the primordial helium mass fraction $Y_p \approx 2(n/p)_{\text{BBN}} / [1 + (n/p)_{\text{BBN}}]$ increases from $Y_p^{\text{std}} \approx 0.247$ to $Y_p \approx 0.46\text{--}0.49$ even for the minimal spin-degeneracy case $g_{n^*} = g_n = 2$ (and reaches $Y_p \gtrsim 0.62$ for $g_{n^*} \ge 4$), in stark contradiction with the observed primordial abundance $Y_p^{\text{obs}} = 0.245 \pm 0.003$.

\section{Concluding remarks}
In this work, we have systematically evaluated the potential role of macroscopic Dicke coherence within the context of the neutron lifetime puzzle. We established that a direct application of superradiance to standard three-body weak beta decay is fundamentally impossible due to the continuous phase space of the final state and the antisymmetrization requirements of the fermionic neutrino fields. Even indirectly, when incorporated into the excited-neutron model of Koch and Hummel,\cite{Koch_2024} Dicke superradiance remains inactive due to the low density of neutrons in terrestrial traps, precluding direct superradiant tests of the hypothesis for $\text{eV}$-scale energy gaps.

Extending our analysis to the early universe at $t \sim 1~\text{s}$ shows that Dicke superradiance still fails to operate despite huge baryon densities ($n_b \sim 10^{22}~\text{cm}^{-3}$) exceeding the laboratory threshold by nearly ten orders of magnitude. Intense $e^\pm$ collisional dephasing destroys inter-nucleon phase coherence, while large thermal photon occupation numbers ($\bar{n}_{\text{th}} \sim 10^6$) enforce detailed balance rather than a one-way superradiant purge. Because an $\text{eV}$-split vacuum state with $\tau_\gamma \sim 100~\text{s}$ inevitably thermalizes prior to weak freeze-out---increasing the effective neutron degrees of freedom and overproducing primordial $^4\text{He}$---BBN rules out such a scenario.

Importantly, this conclusion remains valid across the entire experimentally unconstrained $\Delta E$--$\tau_\gamma$ parameter space ($\Delta E \le 30~\text{keV}$, $\tau_\gamma < 139~\text{s}$), demonstrating that the standard Koch--Hummel scenario is excluded by BBN bounds.

Finally, let us note that although ensembles of fermionic particles, such as free neutrons, can develop collective Dicke-like coherence via their shared radiation fields, there are specific context-dependent subtleties,\cite{Keeling_2014} which we have not treated here, as they remain outside the scope of the present research.

\section*{Acknowledgments}
The author wishes to thank the OpenAI ChatGPT large language model for its assistance in language polishing, manuscript formatting, and LaTeX verification. The author also highly appreciates the insightful comments from the anonymous reviewers, which greatly helped to refine and sharpen the focus of this paper.

\appendix
\section{{Free neutron $\beta$-decay}}
Experimental and theoretical developments in neutron $\beta$-decay, the process in which a neutron is transformed into a proton, an electron, and an electron antineutrino, continue to play a significant role in understanding the electroweak sector of the Standard Model.\cite{Nico_2009,Gonzalez-Alonso_2018,Fomin_2022} Within the Standard Model, neutron $\beta$-decay is caused by the fundamental weak-interaction process: conversion of a down quark into an up quark accompanied by the emission of a virtual $W$ gauge boson. As a result, the corresponding decay width, which can be considered as the reciprocal of the neutron lifetime since $\beta$-decay is its dominant decay channel, contains the Cabibbo--Kobayashi--Maskawa weak mixing matrix element $V_{ud}:$\cite{Fomin_2022}
\begin{equation}
    \Gamma=\frac{1}{\tau}=\frac{fm_e^5c^4}{2\pi^3\hbar^7}\,|V_{ud}|^2\left(|G_V|^2+3|G_A|^2\right),
    \label{A1}
\end{equation}
where $m_e$ is the electron mass, $f$ is a phase-space factor, and $G_A$ and $G_V$ are the axial-vector and vector coupling constants, respectively. The Cabibbo--Kobayashi--Maskawa matrix element $V_{ud}$ can be extracted from experimental information on superallowed $0^+\to 0^+$ nuclear $\beta$-decay transitions.\cite{Hardy_2020}

Equation~\eqref{A1} is the lowest-order result. Radiative corrections are of various types.\cite{Wilkinson_1982,Nico_2009} Namely, they include the final-state electromagnetic interaction between the outgoing proton and electron, corrections from proton recoil, finite nucleon size, and various long-distance and short-distance electroweak and QCD corrections. Including these corrections leads to the Standard Model prediction~\eqref{eq0}, cited in the main text.

\bibliography{NLFTP.bib}
\end{document}